\documentclass[twocolumn,letterpaper,aps,prc,superscriptaddress,nofootinbib,showpacs,floatfix]{revtex4-1}
\usepackage{graphicx}
\usepackage{comment}
\usepackage{setspace}
\usepackage{amsmath}
\usepackage{amssymb}
\usepackage[textwidth=16cm,textheight=22cm]{geometry}
\usepackage{color}
\usepackage{indentfirst}
\usepackage{xspace}
\usepackage{hyperref}
\usepackage{verbatim}
\usepackage{epstopdf}
\usepackage{lineno}

\definecolor{red}{rgb}{1.0, 0.0, 0.0}
\definecolor{orange}{rgb}{1.0, 0.49, 0.0}
\definecolor{orangev2}{rgb}{0.91, 0.45, 0.32}   
\definecolor{blue}{rgb}{0.0, 0.0, 1.00}        
\definecolor{violetv2}{rgb}{0.57, 0.36, 0.51}

\begin{document}
\title{Study of multiplicity dependence of heavy flavor production in  p$-$p collisions  using  rope hadronization mechanism} 

\author{Tulika Tripathy}
\email{tulika.tripathy@cern.ch}
\affiliation{Department of Physics, Indian Institute of Technology Bombay, Mumbai, 40076}	
\author{Bharati Naik}
\email{bharati@phy.iitb.ac.in}
\affiliation{Department of Physics, Indian Institute of Technology Bombay, Mumbai, 40076}	
\author{Ranjit Nayak }
\email{ranjit.nayak@cern.chj}
\affiliation{Department of Physics, Indian Institute of Technology Bombay, Mumbai, 40076}	
\author{Nirbhay~Behera }
\email{nirbhaykumar@cutn.ac.in}
\affiliation{Department of Physics, Schools of Basic and Applied Sciences, Central University of Tamil Nadu, Thiruvarur 610005, India} 
\author{Basanta K. Nandi }
\email{basanta@iitb.ac.in}
\affiliation{Department of Physics, Indian Institute of Technology Bombay, Mumbai, 40076}	
\author{Sadhana Dash }
\email{sadhana@phy.iitb.ac.in}
\affiliation{Department of Physics, Indian Institute of Technology Bombay, Mumbai, 40076}

\begin{abstract}
The multiplicity dependence of the production of the charm mesons in p$-$p collisions at $\sqrt{s} = 7$ TeV and 13 TeV as measured by ALICE experiment  has been  investigated using Pythia 8 event generator by studying the effect of various  processes at partonic level such as the effect of different modes of color reconnections and rope hadronization. The relative yields ($\rm Yield/\langle Yield \rangle$) of D-mesons and  $J/\psi$ as a function of relative charged particle multiplicity for various transverse momentum ($p\textsubscript{T}$) ranges as measured by the ALICE experiment are  in reasonable agreement with the estimations of Pythia 8 model within the framework of microscopic processes. The relative yields of B mesons for various $p\textsubscript{T}$ intervals ($1 <  p_{T}  < 20$ GeV/$c$) have also been predicted in p$-$p collisions at $\sqrt{s} = 7$ TeV and  $\sqrt{s} = 13$ TeV. 
\end{abstract}
\maketitle

\section{Introduction}
\label{Intro}
Heavy quarks like charm and beauty are produced in ultra-relativistic hadronic collisions by hard scatterings between partons of the incoming hadrons. Due to their large mass ($m_{Q} \gg \Lambda_{QCD}$), heavy quarks are produced at the very initial stages of the collisions and the production cross-section is well described by the perturbative Quantum Chromodynamics (pQCD) calculations. The theoretical calculations based on general-mass variable-flavour-number scheme, GM-VFNS \cite{Gmvfns} or fixed order with next-to-leading-log resummation, FONLL \cite{FONLL1} - \cite{FONLL4} predict the inclusive production cross-sections of charm mesons in p$-$p collisions at Large Hadron Collider (LHC). The collinear factorisation theorem at next to leading order (NLO) has been implemented in both the calculations. Alternatively, the leading order (LO) calculations based on the framework of $k_{T}$- factorisation \cite{kTfact} describes the D-meson production cross-section in p$-$p collisions at the LHC.  \\

The production of heavy-flavour particles as a function of charged particle multiplicity in p$-$p collisions at LHC energies is relevant as it allows one  to study the production as a function of event activity at the partonic level as the  charged particle multiplicity is closely associated with the number of multi-partonic interactions (MPIs).  The MPIs refer to many hard and semi-hard partonic interactions occurring in a single collision. As heavy flavors are predominantly produced via hard scattering processes  while charged particle production is dominated by soft processes, this study can  illuminate  about the interplay between hard and soft mechanisms, specifically, the influence of underlying event activities in particle production. 

At LHC energies, the  particle production depends on the beam energy as well as on the impact parameter of the two colliding protons. This can affect the contributions emanating from gluon radiation and multi-partonic interactions (MPI) which can influence the production of heavy flavor quarks. The fluctuations in the gluon density also affects the particle production in high multiplicity regime.  Therefore, the final state particles produced in the collision can be described by  a two-component approach where the hard component is well described by  pQCD inspired models while the soft one depends on phenomenological modelling of underlying events.   

The relative yield of  charm hadron production has been measured as a function of charged particle multiplicity with the ALICE experiment at the LHC in p$-$p collisions at $\sqrt{s}$ = 7 TeV \cite{ALICE}. These heavy flavour particles are measured from the reconstruction of prompt D-mesons and non-prompt J/$\Psi$.  The relative yield of D-meson species ( $D^{0}$, $D^{\pm}$ and $D^{*}$) is observed to increase with an increase in charged particle multiplicity and the trend is similar across all the measured $p_{\rm T}$ intervals. The relative yields of average D-meson species as a function of charged particle multiplicity have faster increase at higher multiplicities and  exhibits a deviation from  linear behaviour.  At central rapidity, the yield of open charm and hidden charm hadrons show similar increase with multiplicity, which indicates that enhancement of the relative yields of heavy flavour is due to c$\bar{\rm c}$ and b$\bar{\rm b}$ production process rather than hadronization. Several model studies such as Pythia 8\cite{pythia8}, EPOS3 \cite{epos1}\cite{epos2}, and percolation calculation \cite{perc1}\cite{perc2} have been performed to describe the open heavy flavour relative yield as a function of charged particle multiplicity. EPOS3 and percolation calculation qualitatively describe the enhancement of the relative yield of open heavy flavour hadrons with charged particle multiplicity. However, Pythia 8 underestimated the data at high multiplicities. 
In heavy-ion collisions, the produced system undergoes a collective expansion (described well by hydrodynamics) and influences the transverse momentum
distributions of light hadrons. The recent measurements in high-multiplicity p$-$p collisions at the LHC  mimics such a
collective behaviour. If heavy flavor quarks participated in such a collective motion in high-multiplicity events, their relative yields might vary as a function of $p_{T}$.  The EPOS3 with hydro component was able to describe the qualitative non-linear evolution of the yield with multiplicity .
In percolation model,  the target and projectile in the high energy hadronic collisions  interact by exchange of  colour sources between them. These colour sources have finite size  and their number is reduced effectively in high density collisions due to coherence.  As a result, the multiplicity due to soft sources decrease while the hard sources are not affected by this. 

In recent studies \cite{RHNew}, it has been observed that  the microscopic model of rope hadronization along with color reconnection mecahnism  implemented in Pythia 8 \cite{RH1}\cite{RH2}, successfully described the enhancement of strange and multi-strange hadrons in p$-$p collisions and it does not assume the formation of a de-confined and thermalized  plasma state. In high multiplicity regime, the colored strings tend to overlap with each other to form colored ropes  with higher effective string tension which eventually hadronizes to particles with higher mass.  The  primary aim of this work is to investigate the  effect of rope hadronization  on the heavy flavour production.  The relative yield of D-meson, B-meson and $J/\psi$ as a function of charged particle multiplicity has been studied with rope hadronization mechanism in p$-$p collisions at  $\sqrt{s}$ = 7 TeV and $\sqrt{s}$ = 13 TeV for three different modes of color reconnection mechanism.

\section{ANALYSIS}  

The analysis is based on 100 million inelastic, non-diffractive events generated with soft-QCD processes using Pythia 8 generator for p$-$p collisions at $\sqrt{s} = 7$ TeV and $\sqrt{s} = 13$ TeV.  Pythia 8 is a Monte-Carlo event generator which has been frequently used in high energy physics, specially for e$-$e, p$-$p and $\mu$-$\mu$ collisions.  Pythia 8 \cite{pythia8} is the successor of Pythia 6 \cite{pythia6},  with some introduction of new physics  processes like multi-partonic interactions (MPI), color reconnections (CR)  etc. and  some improvisations in the existing processes at both partonic and hadronic level. One of the major improvements in Pythia 8 is the involvement of $c$ and $b$ quarks in  MPI $2\rightarrow 2$ hard sub-processes. The details of the physics processes and its implementation can be found in reference \cite{pythiadetails}.\\   
The idea of color reconnections stems from the fact that the multiple interactions  can lead to  generation of many color strings. It is not unreasonable to  consider to connect the strings in appropriate
manner to reduce the string length and hence the potential energy. The way these strings are connected leads to three modes of color reconnections in Pythia 8. 
The multi-parton interactions (MPI) and different modes of color reconnections (CR- 0/1/2) are enabled to study the effect of these mechanism on the heavy flavor yield. Additionally, the effect of rope hadronization (RH) has been studied. More details on RH can be found in \cite{RH3}\cite{RH4}. The various combinations of processes used for this study are: (i) RH on CR(0) on, (ii) RH on CR(1) on, (iii) RH on CR(2) on, (iv) RH off  CR(0) on, (v) RH off CR(1) on, (vi) RH off CR(2) on. 

The yields of D-mesons (D\textsuperscript{0}, D\textsuperscript{+}, D\textsuperscript{*+}), B-mesons (B\textsuperscript{0}, B\textsuperscript{+}) and their charge conjugates are estimated in the mid-rapidity region, $|y| < 0.5$ in five $p\textsubscript{T}$ intervals, from 1 GeV/$c$ to 20 GeV/$c$. The charged particle multiplicity class definition is obtained within the acceptance of ALICE V0 detector ( $-3.7 < y < -1.7$ and $2.8 < y < 5.1$) to reduce the auto-correlation. Thereafter, the charged particle multiplicity for $|\eta| < 0.5$  was obtained for each multiplicity class. The inclusive yields of  $J/\psi$ are  obtained in forward rapidity region, $2.5 < y < 4.0$ for $p\textsubscript{T} > 0$ and the charged particle multiplicity is estimated  within $|\eta| < 1.0$. 
The study was carried out to compare the effect of various processes like color reconnections and rope hadronization  on the yield of heavy flavor  mesons. The obtained  estimations  were compared to the experimental results as measured by ALICE experiment in the same energy\cite{ALICE}. In the experimental result, the relative yield of  D\textsuperscript{0}-mesons are measured  in the mid-rapidity region, $|y| < 0.5$ and the relative charged particles are estimated from the V0 detectors.

\section{RESULTS AND DISCUSSION}  
\label{Results}
\begin{figure}
\begin{center}
\includegraphics[scale=0.4]{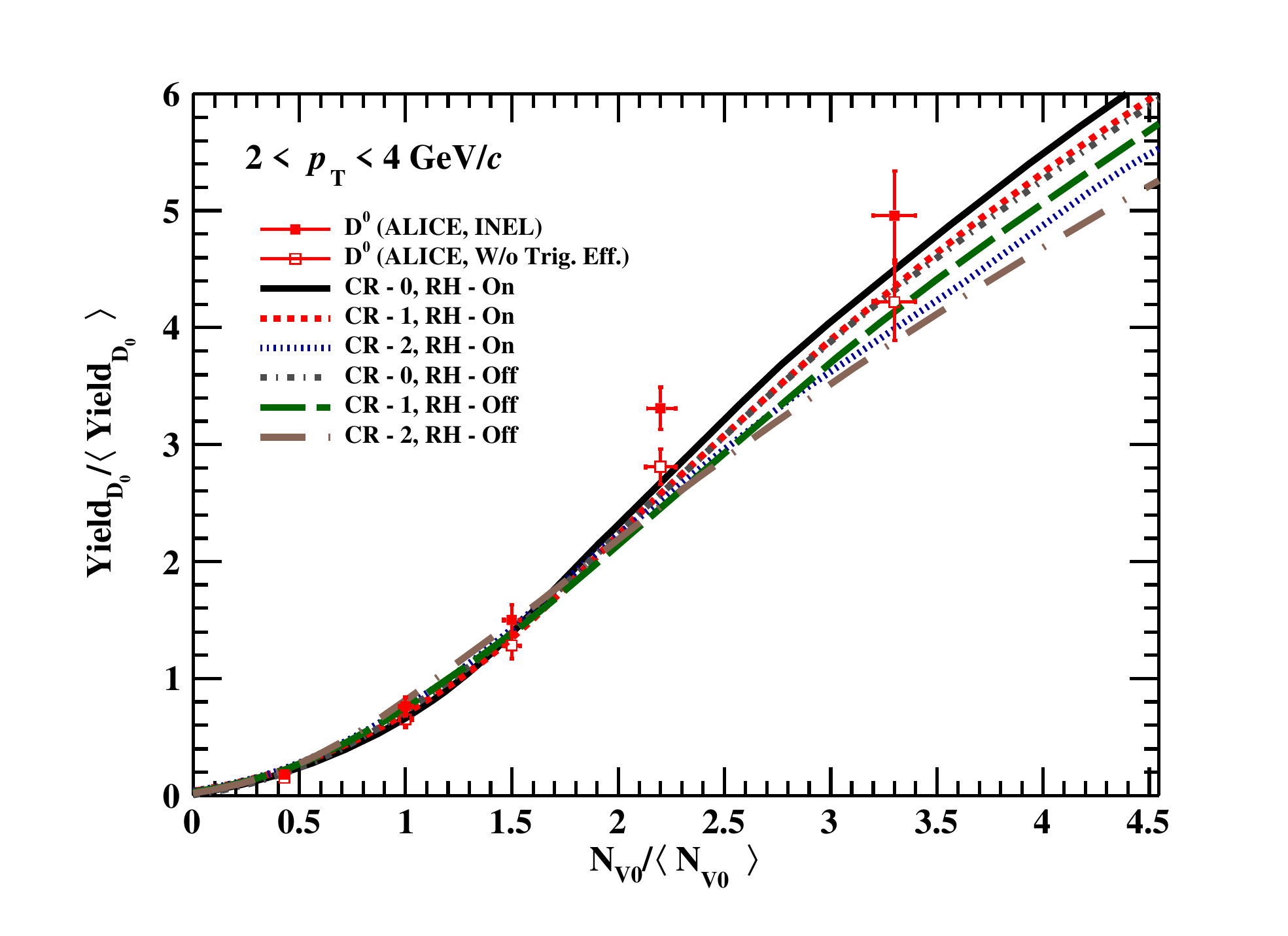}
\includegraphics[scale=0.4]{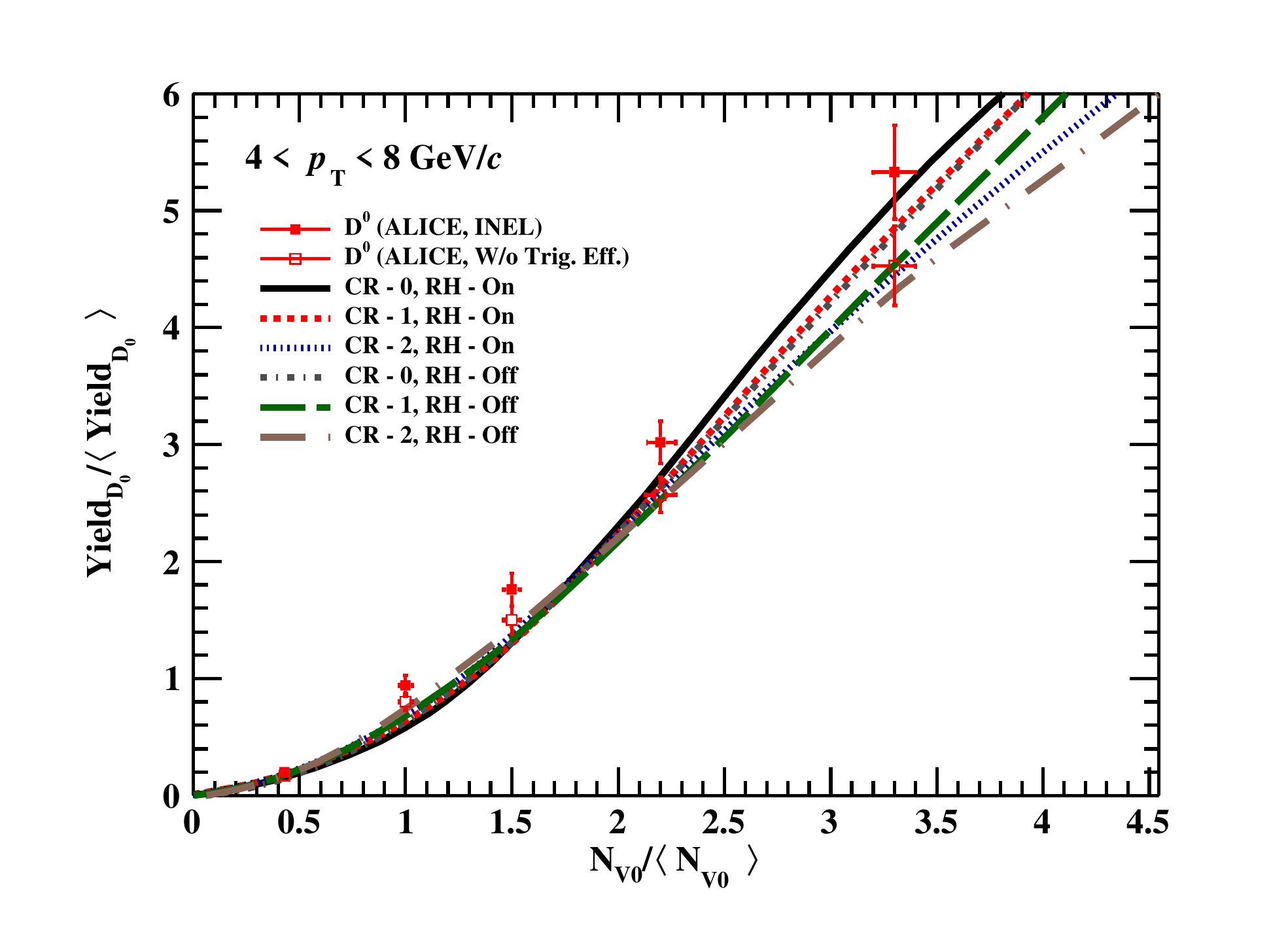}
\caption{Relative D\textsuperscript{0} yield as a function of relative charged-particle multiplicity  in p-$-$p collisions at $\sqrt{s}$ = 7 TeV for the D\textsuperscript{0} $p$\textsubscript{T} range $2 < p\textsubscript{T} < 4$ GeV/$c$ [Upper panel] and $2 < p\textsubscript{T} < 4$ GeV/$c$ [Lower panel].The closed data points measured by ALICE  are normalised to the inelastic cross section. The open data points are not corrected for the trigger selection efficiency and  are normalised to the visible cross section.}
\label{D02to4} 
\end{center}
\end{figure}

The relative yield of D\textsuperscript{0} is obtained as a function of relative charged-particle multiplicity in two $p\textsubscript{T}$ intervals,  $2 < p\textsubscript{T} < 4$ GeV/$c$ and $4 < p\textsubscript{T} < 8$ GeV/$c$  for the various modes of color reconnections available in Pythia 8. The effect of rope formation which is more pronounced in  high multiplicity collisions has also  been considered in this study. The results are  compared to the measured data in Fig.~\ref{D02to4}. The measured  yields by ALICE, represented by the solid squares are normalized  for inelastic cross-section, while the open squares represent  the yields which are not corrected for the trigger selection efficiency factor and are normalised to the visible cross-section.  The relative yields of D\textsuperscript{0}  show an increasing trend with the charged particle multiplicity and the increase is not linear. The formation of color ropes together with color reconnection mechanism qualitatively describes the measured data for all multiplicity classes.  The non-linear rise at higher multiplicity could be due to an increased  production of heavy flavor particles from hadroniziong ropes.  

\begin{figure}
\centering
\includegraphics[scale=0.4]{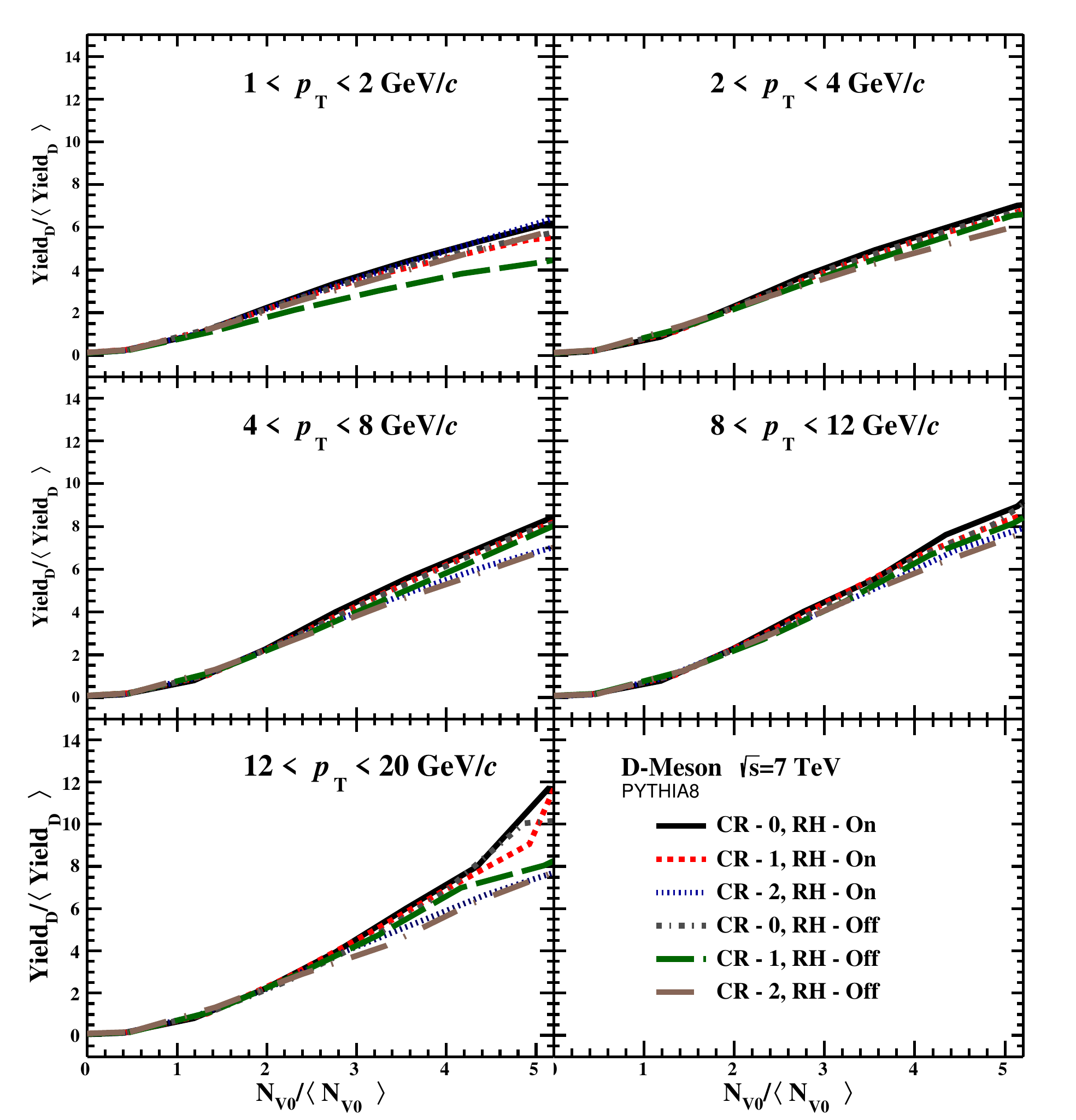}
\caption{The relative D-meson (Avg. of D\textsuperscript{0}, D\textsuperscript{$\pm$}, D\textsuperscript{*$\pm$}) yield as a function of relative charged-particle multiplicity in p$-$p collisions at $\sqrt{s}$ = 7 TeV for different $p$\textsubscript{T} ranges. }
\label{fig:DAvgDiffPt7TeV} 
\end{figure}

\begin{figure}[ht]
\centering
\includegraphics[scale=0.4]{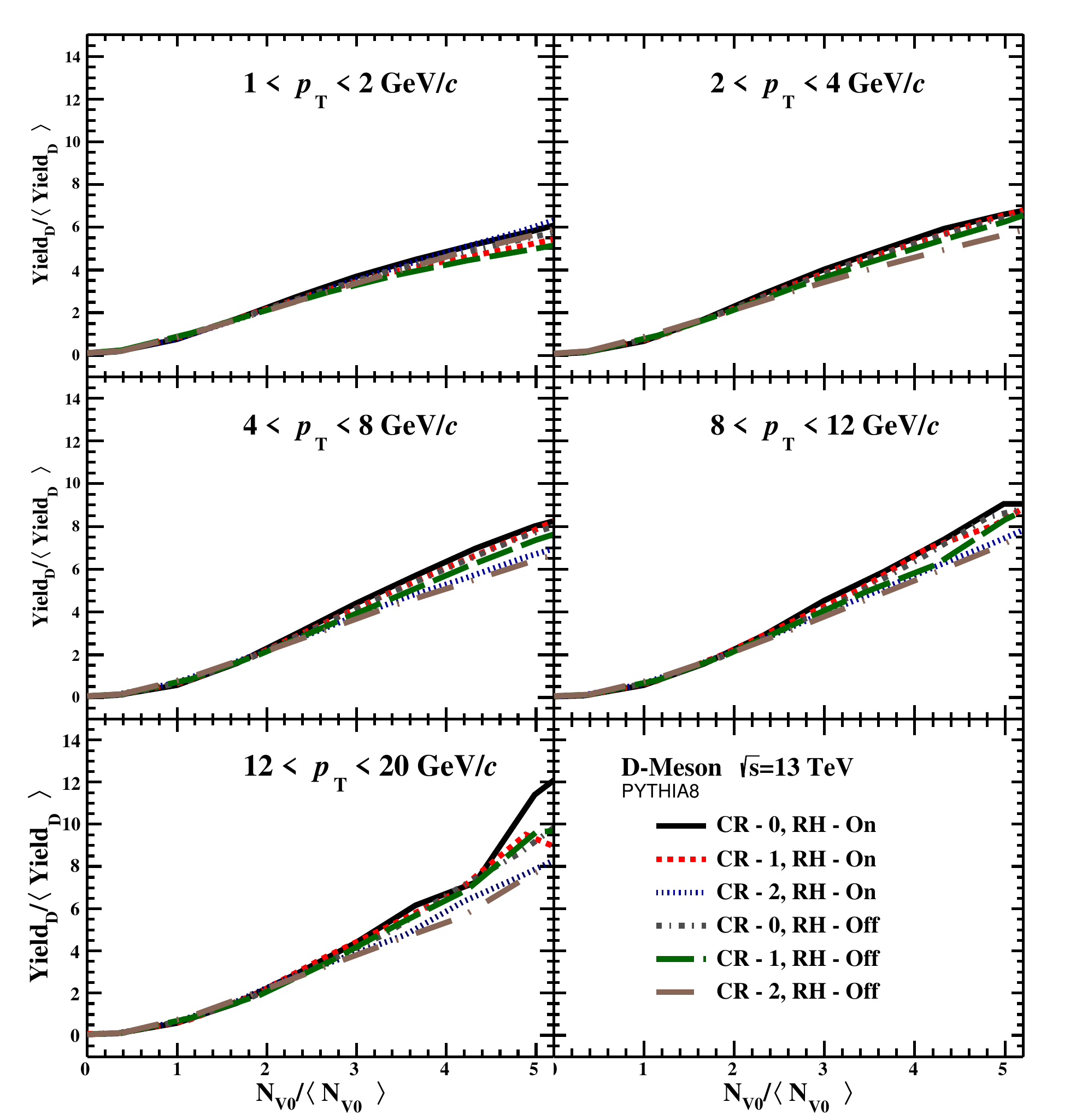}
\caption{Relative D-meson (Avg. of D\textsuperscript{0}, D\textsuperscript{$\pm$}, D\textsuperscript{*$\pm$}) yield as a function of relative charged-particle multiplicity in p$-$p collisions at $\sqrt{s}$ = 13 TeV for different $p$\textsubscript{T} intervals. }
\label{fig:DAvgDiffPt13TeV} 
\end{figure}

The D-meson (mean yield of  D\textsuperscript{0}, D\textsuperscript{$\pm$}, D\textsuperscript{*$\pm$}) relative yields as a function of the relative charged particle multiplicity in p$-$p collisions at $\sqrt{s}$ = 7 TeV  and 13 TeV for five $p\textsubscript{T}$ intervals is shown in Fig.~\ref{fig:DAvgDiffPt7TeV} and Fig.~\ref{fig:DAvgDiffPt13TeV}, respectively. All the shown combinations of different modes of microscopic processes  exhibit  stronger than linear increasing trend with the increase in relative charged particle multiplicity.  A faster increase is observed compared to linear trend for higher multiplicity classes  in different $p\textsubscript{T}$ intervals when the rope hadronization is implemented. The relative yields of D-mesons increase with an increase in $p\textsubscript{T}$ intervals for both the energies. It can be observed form  Fig.~\ref{fig:DAvgDiffPt7TeV} and Fig.~\ref{fig:DAvgDiffPt13TeV} that, there is no considerable collision energy dependence on the relative yields of D-mesons indicating that the charm production is dependent on underlying event activity which is  more pronounced with increasing multiplicity. 

\begin{figure}[ht]
\centering
\includegraphics[scale=0.4]{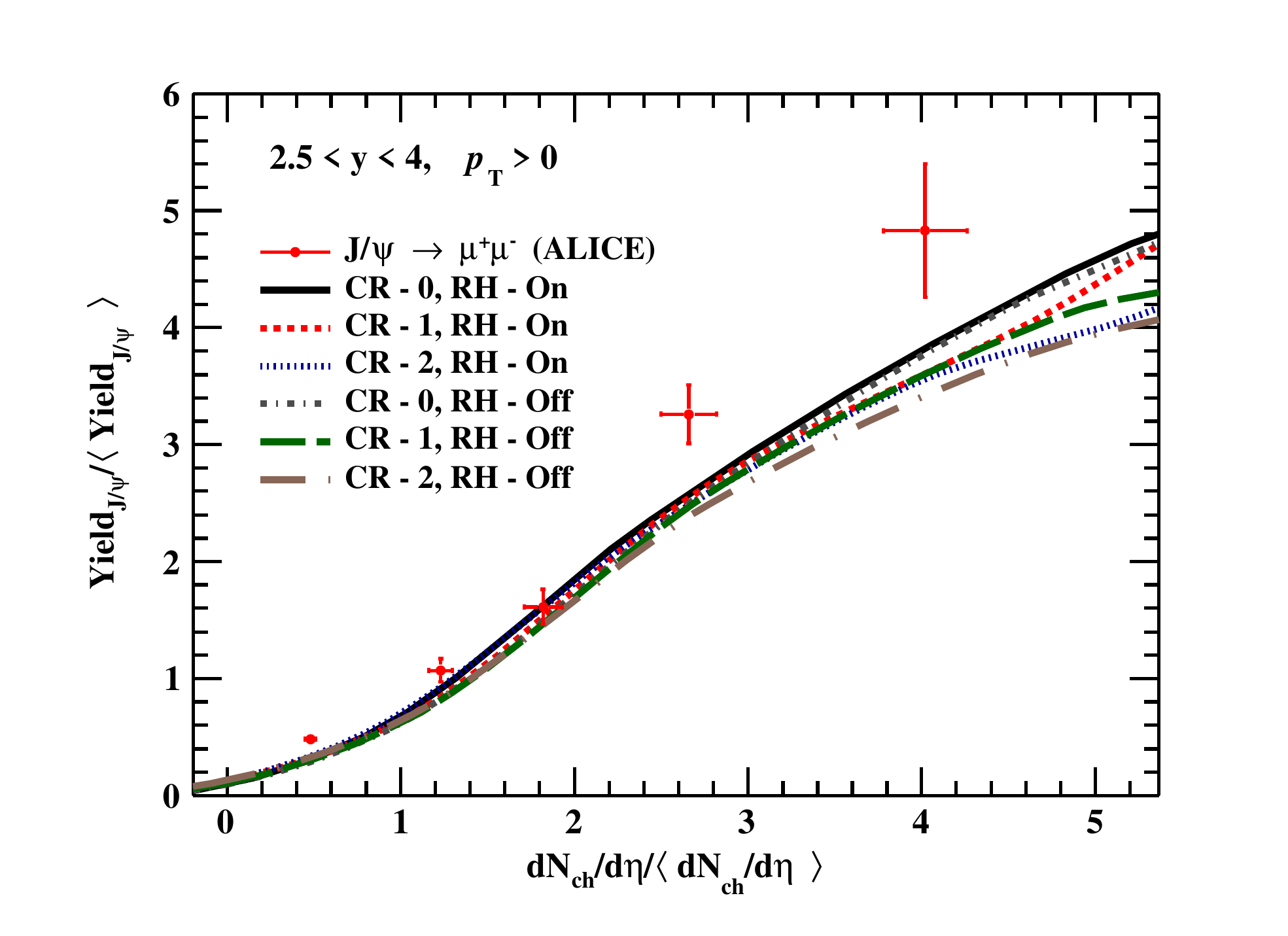}
\caption{The relative yield of inclusive J/$\psi$ as a function of relative charged-particle multiplicity in p$-$p collisions at $\sqrt{s}$ = 7 TeV for $p\textsubscript{T} > 0$. }
\label{fig:InclJpsi} 
\end{figure}

The relative yield of inclusive J/$\psi$ as a function of relative charged particle multiplicity is obtained in Pythia 8 for p$-$p collision at $\sqrt{s} = 7$ TeV. The obtained values are compared with the measured ALICE data and shown in Fig.~\ref{fig:InclJpsi}. It is evident from Fig.~\ref{fig:InclJpsi} that the formation of color ropes along with color reconnection mechanism in Pythia 8 qualitatively describes the increase in relative yield of inclusive J/$\psi$ measured in data. However, it underestimates the measured data. This might suggest that the production of J/$\psi$ mesons are not significantly  affected by underlying event mechanisms.

\begin{figure}[ht]
\centering
\includegraphics[scale=0.4]{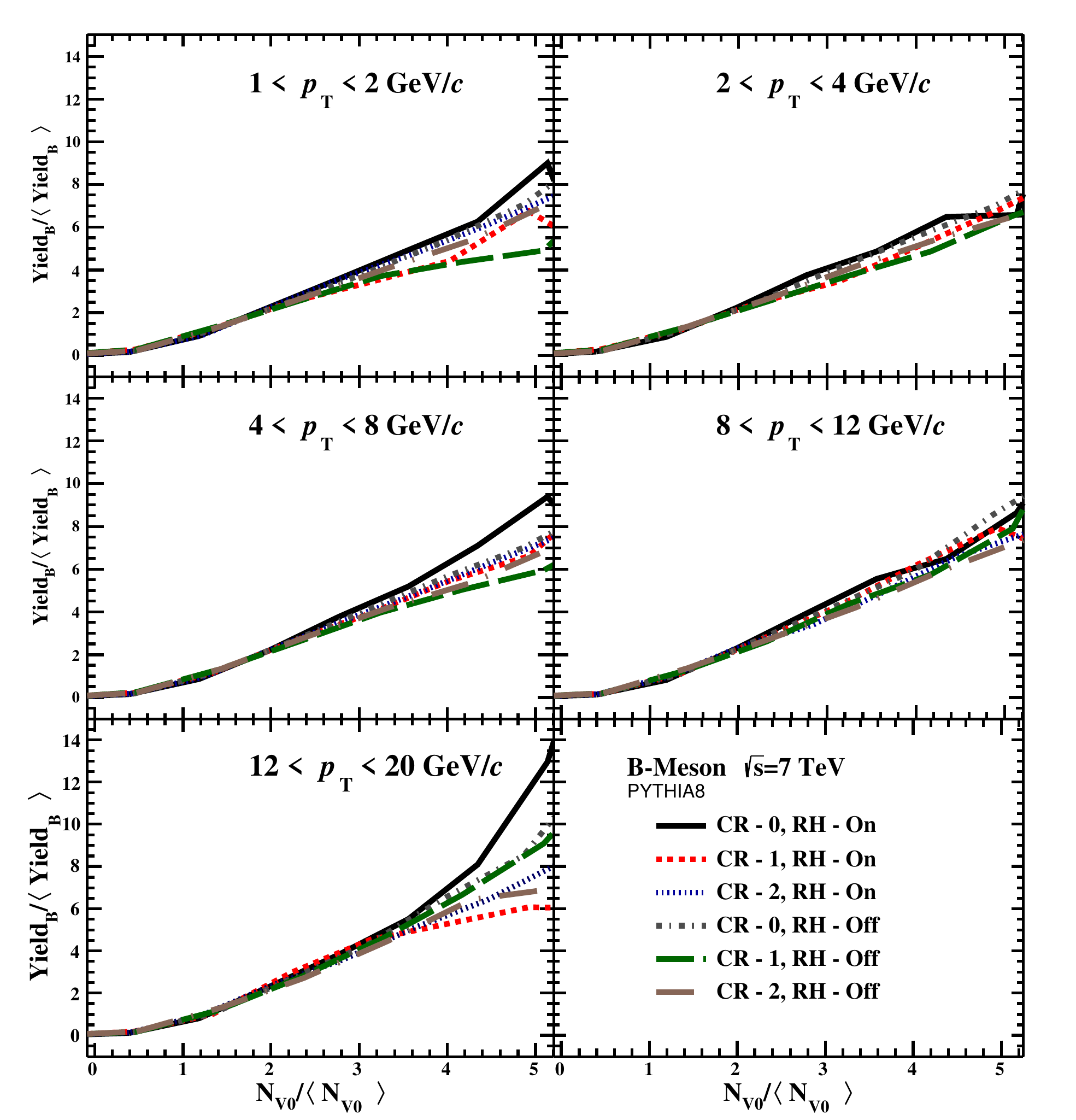}
\caption{Relative B-meson (Avg. of B\textsuperscript{0}, B\textsuperscript{+} and their charge conjugate) yield as a function of relative charged-particle multiplicity in p$-$p collisions at $\sqrt{s}$ = 7 TeV for different $p$\textsubscript{T} intervals. }
\label{fig:BAvgDiffPt7TeV} 
\end{figure}

\begin{figure}[ht]
\centering
\includegraphics[scale=0.4]{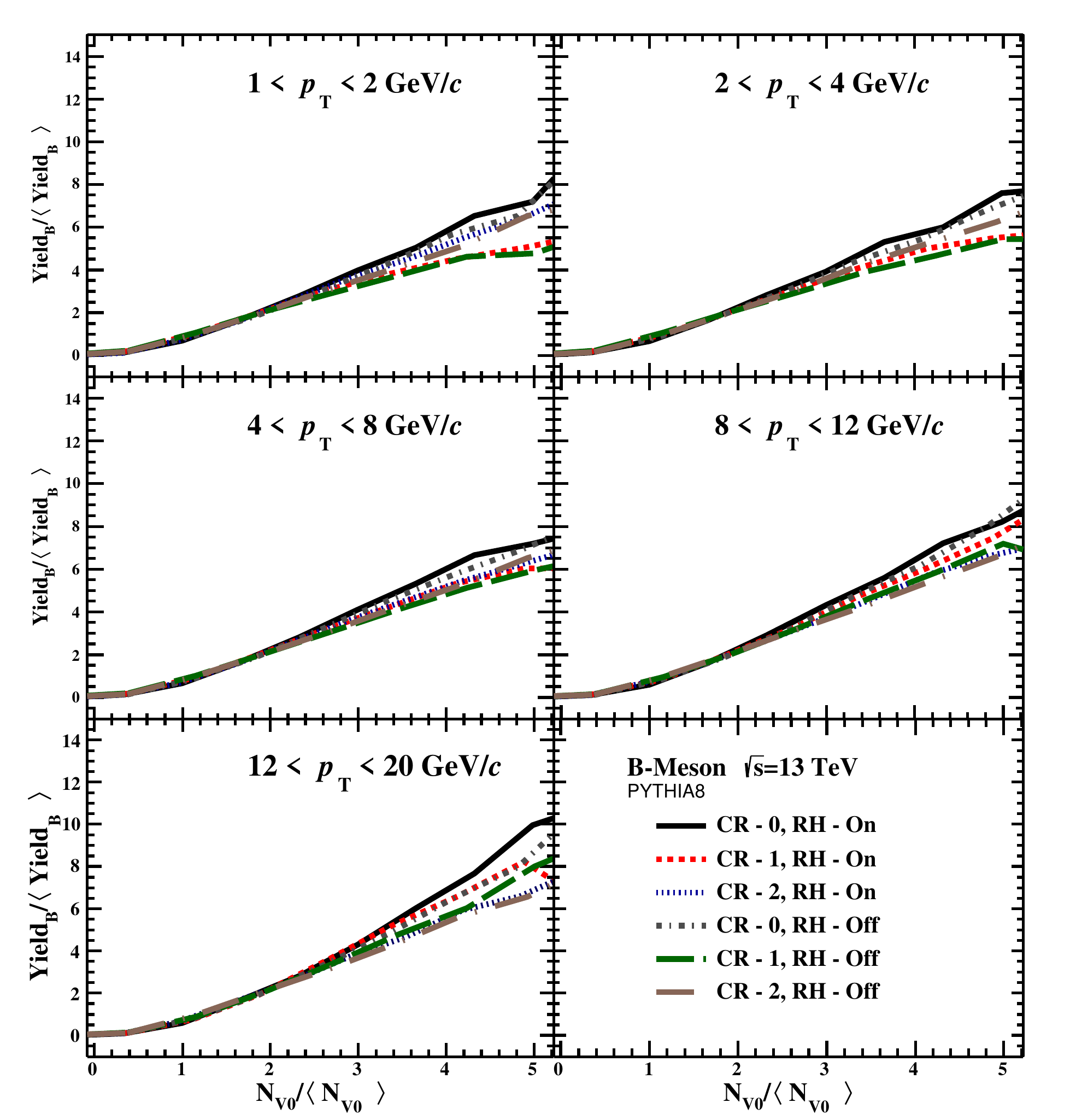}
\caption{Relative B-meson (Avg. of B\textsuperscript{0}, B\textsuperscript{+} and their charge conjugate) yield as a function of relative charged-particle multiplicity in  p$-$p collision at $\sqrt{s}$ = 13 TeV for different $p$\textsubscript{T} intervals. }
\label{fig:BAvgDiffPt13TeV} 
\end{figure}

An additional study was carried out in the beauty sector to estimate the multiplicity dependence of B-mesons. Figure~\ref{fig:BAvgDiffPt7TeV} and  Fig.~\ref{fig:BAvgDiffPt13TeV} depicts the  relative yields of B-meson (Avg. of B\textsuperscript{0}, B\textsuperscript{+} and their charge conjugates) as a function of the relative charged particle multiplicity in p$-$p collisions at $\sqrt{s}$ = 7 TeV  and 13 TeV, respectively, obtained with different  processes  for five $p\textsubscript{T}$ intervals. A similar increasing trend is observed for all the combinations  of Pythia 8 with the relative charged particle multiplicity as seen for D-mesons. It will be an interesting experimental observation in the upcoming measurements at LHC in beauty sector. 
The study describes the multiplicity dependence of D-meson production with rope hadronization mechanism without assuming any collectivity in the produced system while the  J/$\psi$ production is underestimated by the model. A detailed comparison from estimations from other models like Herwig7 and Sherpa can help us understand the contributions emanating from hadronization 
mechansim and other  underlying event processes.

\section{Summary}
\label{Summary}
The multiplicity dependence of the  production of the open charm (D mesons), open beauty (B mesons) and hidden charm (J/$\psi$) mesons  in p$-$p collisions  at  $\sqrt{s} =$ 7 TeV and 13 TeV  has been explored using 
the  different microscopic mechanism of Pythia 8 model  and the estimations have been compared to recent measurements  by ALICE experiment at LHC. The formation of ropes due to overlap of many strings in high multiplicity events along with the partonic color reconnection mechanism provides a  good qualitative and a reasonable quantitative  description of the  multiplicity dependence of  the relative yields of D-mesons and $J/\psi$ in p$-$p collisions at $\sqrt{s} = 7$ TeV measured by the ALICE experiment.  The relative yields of B mesons for various $p\textsubscript{T}$ intervals ($1 <  p\textsubscript{T} < 20$ GeV/$c$) have also been predicted in p$-$p collisions at $\sqrt{s} = 7$ TeV and  $\sqrt{s} = 13$ TeV. The relative yields of D-mesons and B-mesons exhibited a non-linear  increasing trend with an  increase in relative charged particle multiplicity for all $p\textsubscript{T}$ intervals. The observed increasing trend of relative yields is influenced by the $c\bar{c}$ and $b\bar{b}$ production processes. This study can act as an interesting baseline for the upcoming measurements at LHC in heavy flavor sector.

\section{Acknowledgements}
The authors would like to thank the Department of Science and Technology (DST), India for supporting the present work.

\end{document}